\documentclass{UApreprint}
\usepackage{graphics}
\usepackage{fullpage}
\usepackage{verbatim}
\usepackage{amssymb}
\newcommand{\be}{\begin{equation}}
\newcommand{\ee}{\end{equation}}
\newcommand{\bea}{\begin{eqnarray}}
\newcommand{\eea}{\end{eqnarray}}
\newcommand{\bean}{\begin{eqnarray*}}
\newcommand{\eean}{\end{eqnarray*}}
\newcommand{\bi}{\begin{itemize}}
\newcommand{\ei}{\end{itemize}}
\newcommand{\bdm}{\begin{displaymath}}
\newcommand{\edm}{\end{displaymath}}

\newcommand{\ie}{{\it i.e.}}

\newcommand{\lag}{{\mathcal L}}

\newcommand{\meg}{{\mu\to e\gamma}}

\newcommand{\unit}{\mathbf{I}}
\newcommand{\tr}{\mathrm{Tr}}
\newcommand{\gy}{g_\mathrm{Y}}
\newcommand{\Ye}{\mathbf{Y}_e}
\newcommand{\Yn}{\mathbf{Y}_\nu}

\newcommand{\Yu}{\mathbf{Y}_u}
\newcommand{\Ae}{\mathbf{A}_e}
\newcommand{\An}{\mathbf{A}_\nu}

\newcommand{\Au}{\mathbf{A}_u}
\newcommand{\mls}{\mathbf{m}^2_{\tilde{L}}}

\newcommand{\mns}{\mathbf{m}^2_{\tilde{\nu}}}

\newcommand{\mhus}{m^2_{H_u}}

\begin{document}

\title{See-Saw Induced CMSSM Lepton-Flavour Violation Post-WMAP}
\author{Bruce A. Campbell$^1$ David W. Maybury$^1$ and Brandon Murakami$^2$}
\address{$^1$Department of Physics, University of Alberta, Edmonton  AB  T6G 2J1,CANADA \\
$^2$High Energy Physics Division, Argonne National Laboratory, Argonne, IL 60439 USA}

\date{}

\abstract{The see-saw mechanism of neutrino mass generation, when incorporated in supersymmetric theories with supergravity mediated supersymmetry breaking,
results in low-energy lepton-flavour violation arising from the soft supersymmetry breaking slepton masses.
The parameter space of supergravity theories with conserved $R$-parity is severely constrained by the requirement that the
LSP provide cold dark matter with a relic density in the range indicated by the recent WMAP measurements, as well as by laboratory constraints.
We calculate the $\meg$ branching ratio for the constrained minimal supersymmetric standard model, over the range of parameters consistent with WMAP
and laboratory constraints, in families of see-saw model parameterizations which fit the low energy neutrino measurements. We find that over much of
the range of see-saw models, for supersymmetry parameters consistent with WMAP and laboratory bounds,
the resulting predicted rates for $\meg$ (and other charged lepton flavour violating processes)
are within current experimental limits, but that these rates should be detectable
with the next generation of lepton-flavour violation experiments.}

\archive{}
\preprintone{ALTA-TH-12-03}
\preprinttwo{ANL-HEP-PR-03-100}

\submit{}

\maketitle

\section{Introduction}

The standard model contains three continuous global symmetries
associated with lepton flavour. Neglecting non-perturbative effects arising from the weak $\mathrm{SU}(2)$ anomaly, the standard model conserves lepton flavour
exactly, as the charged Yukawa matrix ${\bf Y_e}$ and the gauge interactions can simultaneously be made flavour diagonal.
The solar \cite{Davis}--\cite{SNO}, and atmospheric \cite{SK2},
neutrino deficit observations,
which imply neutrino mass and mixing,
(and their confirmation by reactor \cite{KAMLAND}, and accelerator \cite{K2K}, experiments), presently
provide the only direct observation of physics that cannot be accommodated within the standard model.
The smallness of the inferred neutrino masses can be understood through the see-saw mechanism \cite{see-saw},
which involves the introduction of a heavy Majorana fermion in a gauge singlet (right-handed neutrino) for each generation. The light neutrino masses are then
induced through a Yukawa interaction of the form $N_{i}^c {\bf Y_\nu}^{ij}L_j H$,
once the right-handed neutrinos are integrated out at the
Majorana scale, $M_R$. The resulting induced neutrino mass operator arises at dimension 5 ($HHLL$) and, on dimensional grounds, would be expected to be
the first observable extension beyond the renormalizable dimension 4 operators that compose the standard model interactions, given the standard model particle
content at low energies.

Even without the addition of a see-saw sector generating neutrino masses, the standard model suffers from a gauge hierarchy problem, with quadratic divergences
in radiative corrections to the Higgs mass parameter. These require unnatural fine-tuning of the input Higgs parameters, readjusted at each order in perturbation
theory, in order to maintain the hierarchy of scales between the gravitational energy scale $M_{Pl}$, and the scale of electroweak breaking $M_{W}$. A natural
solution to this problem is the supersymmetric extension of the standard model, where the extra particles and interactions necessitated by supersymmetry
contribute cancelling contributions to the destabilizing quadratic divergences in the Higgs potential, as required by the supersymmetry non-renormalization theorems. After soft supersymmetry breaking the cancellation of the quadratic divergences will still remain, though there will
be finite shifts of the Higgs mass parameters by an amount proportional to the soft supersymmetry breaking. Model dependence enters in the choice of mechanism
to impose the soft supersymmetry breaking. An especially attractive and well-motivated possibility is that the supersymmetry breaking is communicated
(super-)gravitationally from a hidden sector in which supersymmetry is spontaneously broken, to the observable sector of the supersymmetric standard model
(mSUGRA). Models with soft supersymmetry breaking masses of the form that this mechanism would impose, and where each of the soft supersymmetry breaking
scalar masses, gaugino masses and trilinear couplings are universal and flavour diagonal at the Planck scale, comprise the constrained
minimal supersymmetric standard model (CMSSM).

To incorporate see-saw neutrino masses in a supersymmetric extension of the standard model, we consider the minimal
supersymmetric standard model with additional right-handed (singlet) neutrino supermultiplets and their superpotential interactions, where each of the soft supersymmetry breaking scalar masses, gaugino masses and trilinear couplings are universal and flavour diagonal at the Planck scale.
New indirect sources of low-energy lepton-flavour violation (LFV) appear with the introduction of the singlet neutrino supermultiplets. Renormalization group
running of the slepton mass matrices and trilinear couplings in the presence of right-handed neutrinos generates off diagonal elements that contribute to LFV processes \cite{Borzumati:1986},
\be (\Delta m_L^2)_{ij}
\approx - \frac{1}{8\pi^2} (3 + a^2)m_0^2({\bf Y_\nu}^\dagger {\bf Y_\nu})_{ij}
\ln\left(\frac{M_{GUT}}{M_R}\right)
\ee
where $M_R$ is the Majorana scale.
As the see-saw mechanism violates lepton number by two units, the CMSSM with right-handed neutrino singlets
continues to conserve $R$-parity;
therefore the lightest supersymmetric particle (LSP) is stable. If it is assumed that the dark matter is composed of the LSP (which is expected
to be the lightest neutralino), the CMSSM parameter space becomes tightly constrained by the WMAP satellite observations \cite{WMAP}, as well as by laboratory
searches. These constraints \cite{Ellis:2003wm, Ellis:2003bm,susy-WMAP} have important implications for the rates of lepton-flavour violating processes.

In this study, we examine CMSSM lepton-flavour violation \cite{Hisano:1995nq, Hisano:1995cp, susy-LFV} in simple general classes \cite{Casas:2001sr} of see-saw models
which are constructed to fit the low energy neutrino oscillation data. The parametrization of see-saw models is considered in Section 2. Two specific
classes \cite{Casas:2001sr} of them
(corresponding to hierarchical or degenerate Majorana masses for the singlet right-handed neutrinos) form the model range of our calculations in Section 4.
The models considered have their neutrino Yukawa couplings (and Majorana mass scale) chosen as large as reasonable, to maximize the rates for lepton-flavour
violating decays; this is a conservative assumption, as we wish to consider how much of CMSSM parameter space (and see-saw model space) is still consistent
with experimental limits on lepton-flavour violation, and this is conservatively determined under assumptions that maximize its calculated rate.

In Section 3 we discuss the range of parameter space of the CMSSM over which we perform our calculations.
The CMSSM parameters are chosen such that their renormalization group running to low energies yields radiative electroweak symmetry breaking,
and a resulting spectrum of particle masses that is consistent with experiment. In particular, we display our results over CMSSM parameter ranges
determined by \cite{Ellis:2003wm} and \cite{Ellis:2003bm},
which impose that the resulting models have LSP relic densities in the region determined by WMAP \cite{WMAP},
and are consistent with the LEP direct search limits, and the rate for $b \rightarrow s \gamma$. See also \cite{susy-WMAP} for
other parameter determinations using WMAP and laboratory data.

In Section 4 we compute the branching ratio for the decay
$\mu \rightarrow e \gamma$ \cite{Hisano:1995nq,Hisano:1995cp}, in the classes of see-saw models considered, over the allowed range of CMSSM parameter space,
and compare to present \cite{muegam}, and prospective \cite{meg},
data for this process. We consider $\mu \rightarrow e \gamma$ because with the present level of experimental precision, lepton-flavour violation in the models and parameter ranges we consider would only be detectable in muon decays. Of the muon decays, since the rates for  $\mu \rightarrow e e e$, and $\mu N \rightarrow e N$, are largely dominated by electromagnetic penguin contributions (again,
in the models and parameter ranges we consider), they are suppressed with respect to the rate for $\mu \rightarrow e \gamma$ by an extra factor of $\alpha$.
Since at the present time the experimental limit on $\mathrm{BR} (\mu \rightarrow e \gamma) \leq 1.2 \times 10^{-11} $ \cite{muegam}, is of comparable strength to the limits
on $\mathrm{BR} (\mu\rightarrow e e e) \leq 1.0 \times 10^{-12} $ \cite{3e}, and
$\mathrm{BR} (\mu N \rightarrow e N) \leq 6.1\times 10^{-13} $ \cite{mec}, the model class that we study will be consistent
with all the present lepton-flavour violation data if it satisfies the present limit on BR$(\meg)$.
We will find that even for a choice of neutrino Yukawa coupling (and Majorana scale) that maximizes the rates for lepton-flavour violation,
that much of the model space and parameter range is consistent with present experimental limits, though future experiments should probe these
ranges thoroughly, at least for the largest choices of Yukawa couplings.

In Section 5 we present our conclusions. For a thorough review of muon physics and muon flavour violation, see \cite{Kuno:1999jp}.


\section{Supersymmetric See-Saw Parameterization}

The leptonic part of the CMSSM see-saw superpotential is
\be
\lag={\bf Y_e}^{ij} \epsilon_{\alpha \beta} H_d^\alpha e^c_i L^\beta_j +
{\bf Y_\nu}^{ij} \epsilon_{\alpha \beta} H_u^\alpha N^c_{i} L_j^\beta +
\frac{1}{2} {\bf \mathcal{M}}^{ij} N^c_{i} N^c_{j}
\ee
above the Majorana scale. Here, $L_i$, $ i = e,\mu,\tau$, is the left handed
weak doublet, $e^c_i$ is the charged lepton weak singlet, and
$H_u$ and $H_d$ are the two Higgs doublets of opposite hypercharge.
The anti-symmetric SU(2) tensor is defined by $\epsilon_{12}
=+1$. $N$ denotes the right-handed neutrino singlet. The
Yukawa matrices ${\bf Y_e}$ and ${\bf Y_\nu}$ give masses to the charged
leptons and Dirac masses to the neutrinos respectively. The
Majorana matrix, ${\bf \mathcal{M}}^{ij}$, gives the right-handed neutrinos their heavy Majorana mass.
Below the Majorana scale the
right-handed neutrinos are integrated out, and after renormalization down to the scale of electroweak symmetry breaking,
induce a Majorana mass for the light left-handed neutrinos via the see-saw mechanism,
\be
{\bf m_\nu} ={\bf Y_\nu}^T {\bf \mathcal{M}}^{-1} {\bf Y_\nu} <H^0_u>^2
\ee
where $<H^0_u>^2 = v^2_2 = v^2 \sin^2 \beta$ and $ v= (174 \hspace{1mm} \mathrm{GeV})^2$ as set by the
Fermi constant $G_F$. By transforming to a basis where ${\bf Y_e}$
and the gauge interactions are flavour diagonal, the left-handed
neutrino mass matrix is diagonalized by the MNS matrix ${\bf U}$,
\be
{\bf U}^T {\bf m_\nu} {\bf U}= \mathrm{diag}(m_1, m_2, m_3)
\ee
where ${\bf U}$ is a unitary matrix that connects flavour states to the mass eigenbasis.
It is possible to parameterize the MNS matrix as follows,
\be
{\bf U} = {\bf U}^\prime
\mathrm{diag} (e^{-i\phi/2}, e^{-i\phi^\prime},1)
\ee
\be
{\bf U}^\prime= \left( \begin{array}{ccc} c_{13}c_{12}& c_{13}s_{12}& s_{13}e^{-i\delta} \\
-c_{23}s_{12} - s_{23}s_{13}c_{12}e^{i\delta}& c_{23}c_{12} - s_{23}s_{13}s_{12}e^{i\delta} & s_{23}c_{13}\\
s_{23}s_{12} - c_{23}s_{13}c_{12}e^{i\delta} & -s_{23}c_{12} - c_{23}s_{13}s_{12}e^{i\delta} & c_{23}c_{13}
\end{array} \right).
\ee
where $\phi$ and $\phi^\prime$ are additional CP violating
phases and ${\bf U}^\prime$ has the usual form of the CKM matrix. It was
shown in \cite{Casas:2001sr} that the Yukawa matrix $Y_\nu$ can be
re-expressed in a simple general form. By defining,
\be
{\bf \kappa} \equiv
\frac{{\bf m_\nu}}{<H^0_u>^2} = {\bf Y_\nu}^T \mathcal{M}^{-1} {\bf Y_\nu}
\ee
and using the MNS matrix, it is possible to diagonalize $\kappa$,
\be
{\bf \kappa}_d = {\bf U}^T {\bf Y_\nu}^T {\bf \mathcal{M}}^{-1} {\bf Y_\nu} {\bf U}.
\ee
where the $d$ subscript denotes diagonalization. It is always possible to make an arbitrary field re-definition to rotate to a
basis such that ${\bf \mathcal{M}}$ is diagonal, hence ${\bf \mathcal{M}}_d =
\mathrm{diag}(\mathcal{M}_1,\mathcal{M}_2 \mathcal{M}_3)$. In this case,
\be
{\bf 1} = \left(\sqrt{{\bf \mathcal{M}}_d^{-1}} {\bf Y_\nu} {\bf U} \sqrt{{\bf\kappa}_d^{-1}}\right)^T
\left(\sqrt{{\bf\mathcal{M}}_d^{-1}} {\bf Y_\nu} {\bf U} \sqrt{{\bf\kappa}_d^{-1}}\right)
\ee
where a square root over a diagonal matrix denotes the positive square root of its entries.
One then identifies
\be
{\bf R} \equiv \sqrt{{\bf\mathcal{M}}_d^{-1}} {\bf Y_\nu} {\bf U} \sqrt{{\bf\kappa}_d^{-1}}
\ee
as an arbitrary orthogonal matrix. Then, the most general form of ${\bf Y_\nu}$ is \cite{Casas:2001sr}
\be
{\bf Y_\nu} = {\sqrt{{\bf \mathcal{M}}_d}} {\bf R} \sqrt{{\bf\kappa}_d} {\bf U}^\dagger.
\label{ynu}
\ee
As pointed out by the authors of \cite{Casas:2001sr}, the physical
low-energy observables contained in ${\bf U}$ and ${\bf \kappa}_d$
are augmented by three positive mass eigenvalues associated with
${\bf \mathcal{M}}$ and three (in general complex) parameters that
define the orthogonal matrix ${\bf R}$. It should be stressed that the
above equation is defined at the Majorana scale, $M_R$.
It is useful to parameterize the neutrino Yukawa couplings with
the use of an arbitrary orthogonal matrix, ${\bf R}$, as it allows a
general examination of the origin of flavour violation in see-saw models.

Following \cite {Casas:2001sr} we will consider two classes of neutrino hierarchy models.
In the first case we will examine a strong right-handed neutrino hierarchy and in the second, we will consider degenerate right-handed neutrinos.
In both cases we will assume that the Yukawa couplings of the left-handed neutrinos are hierarchical. We will impose the
condition that the largest eigenvalue of the ${\bf Y_\nu}^\dagger {\bf Y_\nu}$ matrix
(denoted $|Y_{0}|^{2}$) coincide with the square of the top quark Yukawa coupling $|Y_{t}|^{2}$ at the unification scale
$M_{\mathrm{GUT}}$. This Yukawa unification condition is suggested in simple $SO(10)$ models, and has the effect of making the neutrino Yukawa couplings,
and hence the rates for lepton-flavour violating processes, as large as reasonably possible. Since we are interested in the degree to which
present experimental
limits rule out regions of model and CMSSM parameter space, maximizing the expected rates gives us a conservative determination of the models
and parameter ranges that are still viable.
More specific details of the classes of models to be analyzed will be discussed in Section 4.


\section{Supersymmetry Breaking and the CMSSM}

Since supersymmetric particles have not yet been observed, the
model Lagrangian must contain terms that break
supersymmetry. If we assume that supersymmetry is broken softly, in
that the supersymmetry violating terms are of mass dimension 2 and
3, then the Lagrangian has the following supersymmetry breaking terms,
\bea
-\mathcal{L}_{soft} &=& ({\bf m^2_{\tilde L}})_{ij}\tilde L^\dagger_i
\tilde L_j + ({\bf m^2_{\tilde e}})_{ij}\tilde e^*_{Ri} \tilde e_{jR} +
({\bf m^2_{\tilde \nu}})_{ij} \tilde
\nu^*_{Ri} \tilde \nu_{Rj}\nonumber \\
&& +({\bf m^2_{\tilde Q}})_{ij}\tilde Q^\dagger_i \tilde Q_j + ({\bf m^2_{\tilde u}})_{ij}\tilde u^*_{Ri} \tilde u_{jR} + ({\bf m^2_{\tilde d}})_{ij}\tilde d^*_{Ri} \tilde d_{jR}
\nonumber \\
&& + \tilde m_{H_d}^2 H_d^\dagger H_d + \tilde m_{H_u}^2 H_u^\dagger H_u + (B \mu H_d H_u + \frac{1}{2} B_{\nu} {\bf \mathcal{M}}_{ij} \tilde \nu^*_{Ri} \tilde \nu^*_{Rj}
+\mathrm{h.c.}) \nonumber \\
&&[({\bf A_d})_{ij} H_d \tilde d^*_{Ri} \tilde Q_j + ({\bf A_ u})_{ij} H_u \tilde u_{Ri}^* \tilde Q_j + ({\bf A_l})_{ij}H_d \tilde e^*_{Ri}\tilde L_j
+ ({\bf A_\nu})_{ij}H_u \tilde \nu^*_{Ri}L_j
\nonumber \\
&& + \frac{1}{2}M_1 \tilde B^0_L \tilde B^0_L + \frac{1}{2} M_2
\tilde W^a_L \tilde W^a_L + \frac{1}{2} M_3 \tilde G^a \tilde G^a
+ \mathrm{h.c.}]
\label{softy}
\eea
Note the presence of terms containing $\bf{m_{\tilde\nu}}^2$
and $\bf{A}_\nu$ in eq.(\ref{softy}). These terms are only included above the Majorana
scale. Below the Majorana scale, the soft part of the Lagrangian
returns to that of the CMSSM. In the CMSSM scenario, supersymmetry
is broken in a universal $\ie$, flavour independent, manner giving the
following relations
\be
({\bf m}_{\tilde f}^2)_{ij} = m_0^2 {\bf 1}
\hspace{4mm} \tilde m_{h_i}^2 = m_0^2 \hspace{4mm} {\bf A}_{{\bf f} ij} = am_0 {\bf Y_f},
\ee
where $m_0$ is a universal scalar mass and $a$ is a
dimensionless constant. We restrict to the CMSSM in our studies and set the trilinear A-term soft parameter $a=0$.
The ranges for the other non-zero, Planck-scale, inputs to the CMSSM are chosen such that their renormalization group running
to low energies yields radiative electroweak symmetry breaking, and a resulting spectrum of particle masses that is consistent
with experiment. In particular, we display our results over CMSSM parameter ranges determined by
\cite{Ellis:2003wm} and \cite{Ellis:2003bm}, which not only impose that the resulting model have LSP relic densities in the
range determined by WMAP \cite{WMAP}, but that they have spectra consistent with the LEP direct search limits, as well as the
rate for $b \rightarrow s \gamma$. Following these authors we ignore the focus point region in parameter space which occurs at very
large $m_{0}$ and whose location depends on $m_{t}$ and $M_{H}$ in an extremely sensitive manner.

Note that the absence of off-diagonal
terms leads to flavour conservation (up to effects of light neutrino mass splittings).
However, these relations are
imposed at the GUT scale and are therefore subject to
renormalization group running.
Above the Majorana scale, the neutrino sector modifies the CMSSM
renormalization group equations (RGEs). In fact, the
flavour violation is controlled by the off-diagonal terms in
${\bf Y_\nu}^\dagger {\bf Y_\nu}$ which contribute to the off-diagonal terms
of ${\bf m^2_{\tilde L}}$. In the leading log approximation to the RGEs we have,
\bea
({\bf m_{\tilde L}}^2)_{ij} &\approx& - \frac{1}{8\pi^2} (3 + a^2)m_0^2({\bf Y_\nu}^\dagger {\bf Y_\nu})_{ij} \ln \frac{M_{GUT}}{{M_R}} \nonumber \\
({\bf m_{\tilde e}}^2)_{ij} &\approx& 0 \nonumber \\
({\bf{A_e}})_{ij} &\approx& - \frac{3}{8\pi^2} a m_0 Y_{l_i} ({\bf Y_\nu}^\dagger
{\bf Y_\nu})_{ij} \ln \frac{M_{GUT}}{M_R}
\eea
where $Y_{l_i}$ denotes the Yukawa coupling of the of the charged lepton $l_i$.
It is the presence of such terms that leads to significant flavour violation.
We will see in the following sections how much flavour violation we should
expect, and how the branching ratio for the process $\mu \rightarrow e \gamma$ is affected.
The branching ratio for $\mu \rightarrow e \gamma$ can be estimated through mass insertion techniques \cite{Hisano:1995nq,Hisano:1995cp}:
\bea
\mathrm{BR}(\mu \rightarrow e \gamma) &\sim& \frac{\alpha^3}{G_F^2} \frac{(m_{{\bf L}})_{12}^2}{m_s^8} \tan^2 \beta \nonumber \\
&\sim& \frac{\alpha^3}{G_F^2 m_s^8} \left| \frac{-1}{8\pi^2}(3+ a^2)m_0^2\ln\frac{M_{GUT}}{M_R}\right|^2 \left|\left({\bf Y_\nu}^\dagger
{\bf Y_\nu}\right)_{12}\right|^2 \tan^2 \beta
\label{tanbdepend}
\eea
where $m_s$ is a typical slepton mass. We note that the branching ratio is proportional to $\tan^2 \beta$, which will
give an increasing dependence on the ratio of Higgs vevs $\tan \beta$, and will be evident in our detailed results in the next section.


\section{$\mu \rightarrow e \gamma$ In The CMSSM See-Saw}

Following \cite {Casas:2001sr} we consider two classes of neutrino hierarchy models.
In both cases the neutrino Yukawa couplings to the left-handed neutrinos are assumed to be hierarchical.
In the first class of models the Majorana mass terms for the singlet right-handed
see-saw neutrinos are assumed to be
strongly hierarchical.
In the second, we will assume that the right-handed singlet neutrinos have degenerate Majorana masses.
We numerically integrate the one loop CMSSM RGEs with right-handed neutrino supermultiplets.
In addition, we have re-derived the expressions \cite{Hisano:1995nq} for the amplitude for $\meg$,
and we use the resulting full expressions (see Appendix) to calculate the branching ratios.

\subsection{{Hierarchical $\nu_R$s}}

As we saw in Section 2, ${\bf Y_\nu}$ can be expressed using an
orthogonal matrix, ${\bf R}$. Following \cite{Casas:2001sr}, and ignoring possible phases, it is useful to
parameterize R as,
\be
{\bf R} = \left( \begin{array}{ccc} c_{2}c_{3}&-c_1s_3-s_1s_2c_3 & s_1s_3-c_1s_2c_3 \\
c_2s_3& c_1c_3-s_1s_2s_3& -s_1c_3-c_1s_2s_3\\
s_2& s_1c_2& c_1c_2
\end{array} \right).
\label{Rmatrix}
\ee
Since we are assuming that the left-handed neutrinos are hierarchical, we take
\be
\kappa_2 = \sqrt\frac{(\Delta m_\nu^2)_{\mathrm{sol}}}{v^4_2} \hspace{4mm}
\kappa_3 = \sqrt\frac{(\Delta m_\nu^2)_{\mathrm{atm}}}{v^4_2},
\ee
and based on the bi-maximal LMA mixing solution we take the MNS matrix to be,
\be
{\bf U_{MNS}} \approx \left( \begin{array}{ccc} .866&.500&0 \\
-.354&.612&.707\\
.354&-.612&.707
\end{array} \right).
\ee
If we assume a strong hierarchy in the right-handed sector, then
\be
({\bf Y_\nu})_{ij} = \sqrt{\mathcal{M}_3}
\delta_{i3}{\bf R}_{3l} (\sqrt{{\bf \kappa}_d})_l {\bf U}^\dagger_{lj}.
\label{egn}
\ee
The largest eigenvalue of eq.(\ref{egn}) is $Y_0 =
\mathcal{M}_3(|{\bf R}_{32}|^2\kappa_2 + |{\bf R}_{33}|^2\kappa_3)$. We identify
this with the top coupling at $M_{GUT}$ as in the case of many
$SO(10)$ models. By identifying the
largest Yukawa in the neutrino sector with the top coupling LFV
is maximized. We assume that ${\bf R}_{32} \neq 0$ or ${\bf R}_{33} \neq 0$. The pathology of
the case where ${\bf R}_{32} = 0$ or ${\bf R}_{33} = 0$ is discussed in \cite{Casas:2001sr}, which forms
a small region of parameter space.
With these assumptions, $M_R \sim 10^{15}$ GeV.
This leaves us with one complex parameter. Following \cite{Casas:2001sr}, we will assume that this parameter is real.
Therefore,
${\bf Y_\nu}$ will depend on one angle, denoted by $\theta_1$ in eq.(\ref{Rmatrix}).

\begin{figure}[ht!]
   \newlength{\picwidthd}
   \setlength{\picwidthd}{5in}
   \begin{center}
       \resizebox{\picwidthd}{!}{\includegraphics{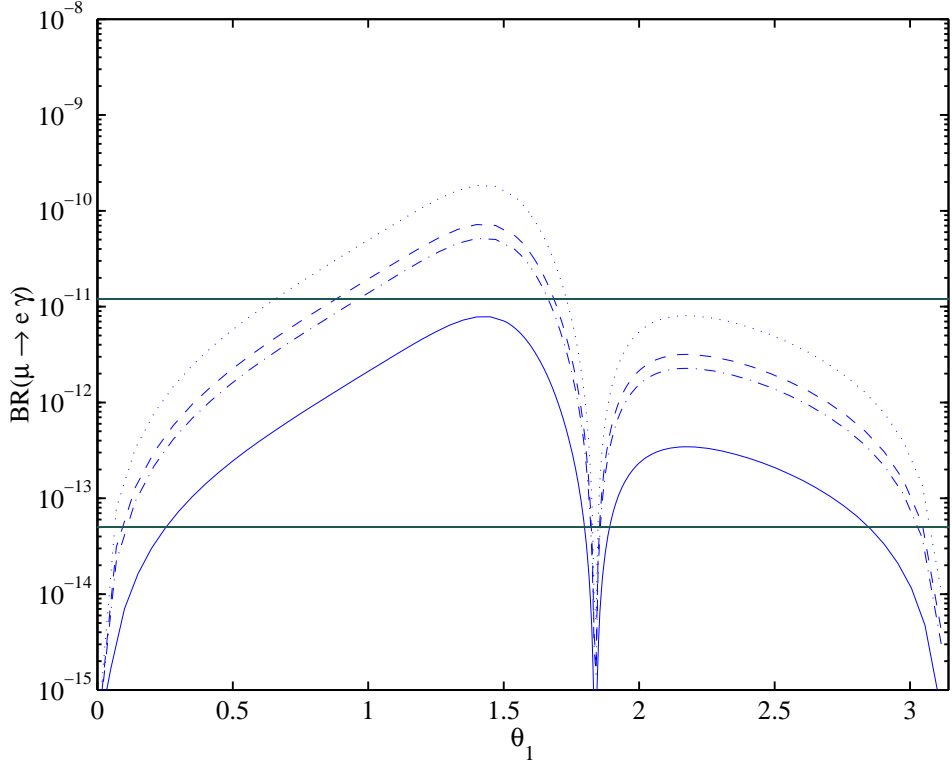}}
   \end{center}
   \caption{{\bf Hierarchical $\nu_R$s}: BR$(\meg)$ as a function of the seesaw parameter $\theta_1$; $\mu>0$ and $a=0$.
   The solid curve corresponds to $\tan \beta = 5$, $m_0 = 140$ GeV, $m_{1/2} = 700$ GeV.
   The dash-dot curve corresponds to $\tan \beta = 10$, $m_0 = 125$ GeV, $m_{1/2} = 560$ GeV.
   The dashed curve corresponds to $\tan \beta = 20$, $m_0 = 200$ GeV, $m_{1/2} = 760$ GeV.
   The dotted curve corresponds to $\tan \beta = 40$, $m_0 = 390$ GeV, $m_{1/2} = 900$ GeV.
   Each parameter set is chosen to lie inside the CMSSM allowed region \cite{Ellis:2003wm}.
   The upper horizontal line indicates the present experimental bound and the lower line
   indicates the expected upcoming experimental sensitivity from MEG.}
   \label{LFV1}
\end{figure}

First, consider figure \ref{LFV1}. This plot shows $\mathrm{BR}(\meg)$ as a function of $\theta_1$ and
is made with parameters typical of the WMAP regions of \cite{Ellis:2003wm}, as indicated in the caption.
The angle $\theta_1$ varies over $0$ to $\pi$ and we only show the $\mu>0$ case as
the plots with $\mu<0$ are very similar.
Most of $\theta_1$ is allowed for low to moderate $\tan \beta \lesssim 40$.
Notice that there are two special places where
the branching ratio becomes highly suppressed. These choices for $\theta_1$
correspond to the vanishing of the off diagonal element $({\bf Y_\nu}^\dagger {\bf Y_\nu})_{12}$
which results in large flavour suppression.
The special angles are,
\be
\tan \theta_1 \approx -\sqrt{\frac{\kappa_3}{\kappa_2}} \frac{{\bf U}^*_{13}}{{\bf U}^*_{12}} \approx 0
\hspace{5mm} \tan \theta_1 \approx -\sqrt{\frac{\kappa_3}{\kappa_2}} \frac{{\bf U}^*_{23}}{{\bf U}^*_{22}}.
\label{sangle}
\ee

\begin{figure}[ht!]
   \newlength{\picwidtha}
   \setlength{\picwidtha}{7in}
   \begin{center}
       \resizebox{\picwidtha}{!}{\includegraphics{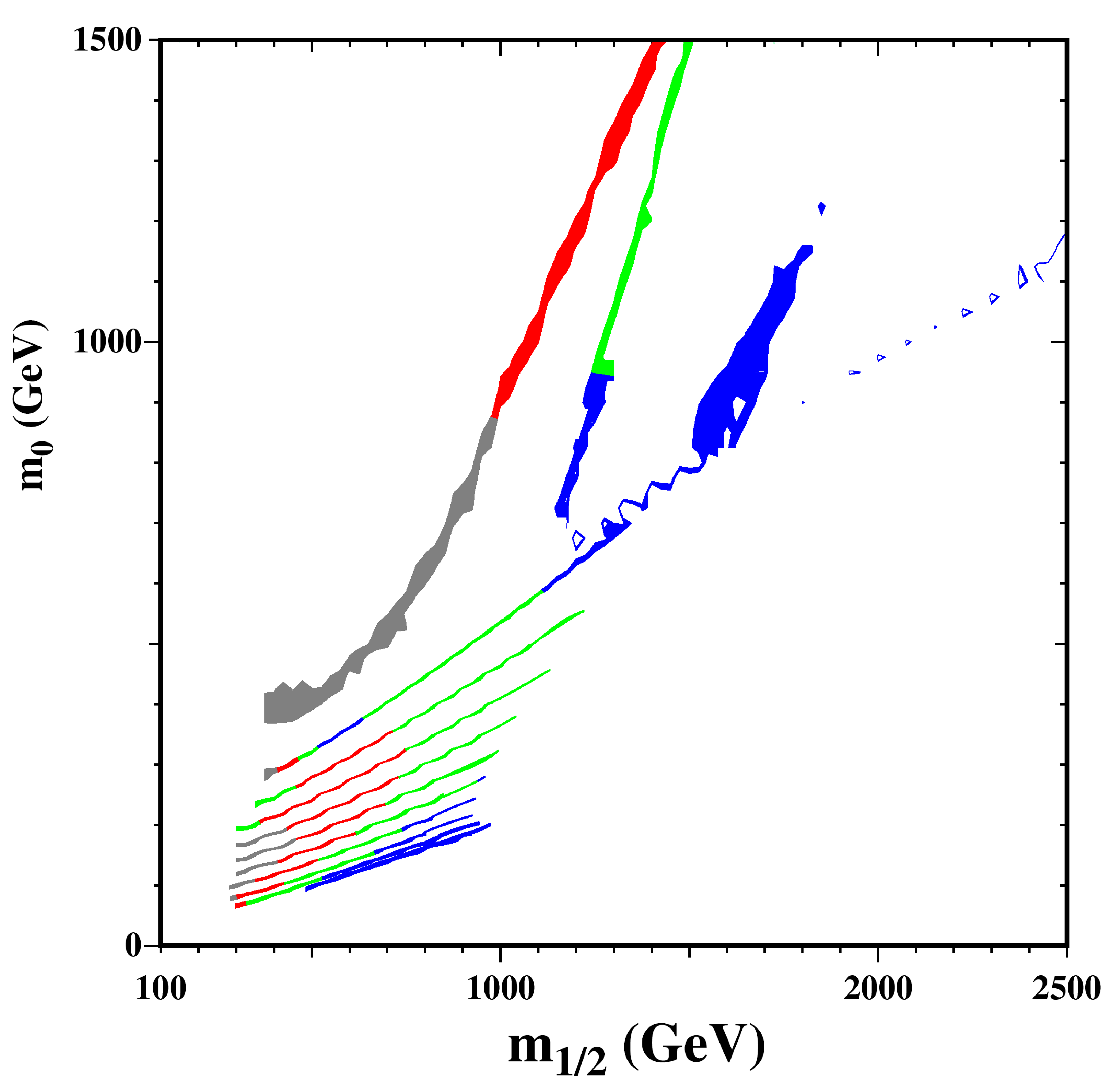}}
   \end{center}
   \caption{{\bf Hierarchical $\nu_R$s}: WMAP and laboratory constraint parameterization of the CMSSM,
   and LFV compliance, based on the current LFV bound, BR$(\meg)<1.2\times 10^{-11}$
   for $\tan \beta =5,10,15,20,25,30,35,40,45,50,55$, $\mu>0$ and $a=0$.
   Grey indicates that less than $25$\% of the range of $\theta_1$ is allowed.
   Red indicates that between $25$\% and $50$\% of $\theta_1$ is allowed.
   Green and blue illustrate that $50$\% to $75$\%, and $75$\% to $100$\%, are allowed respectively.
   The constraint regions are reproduced from \cite{Ellis:2003wm}.}
   \label{WMAP1}
\end{figure}

\begin{figure}[ht!]
   \newlength{\picwidthb}
   \setlength{\picwidthb}{7in}
   \begin{center}
       \resizebox{\picwidthb}{!}{\includegraphics{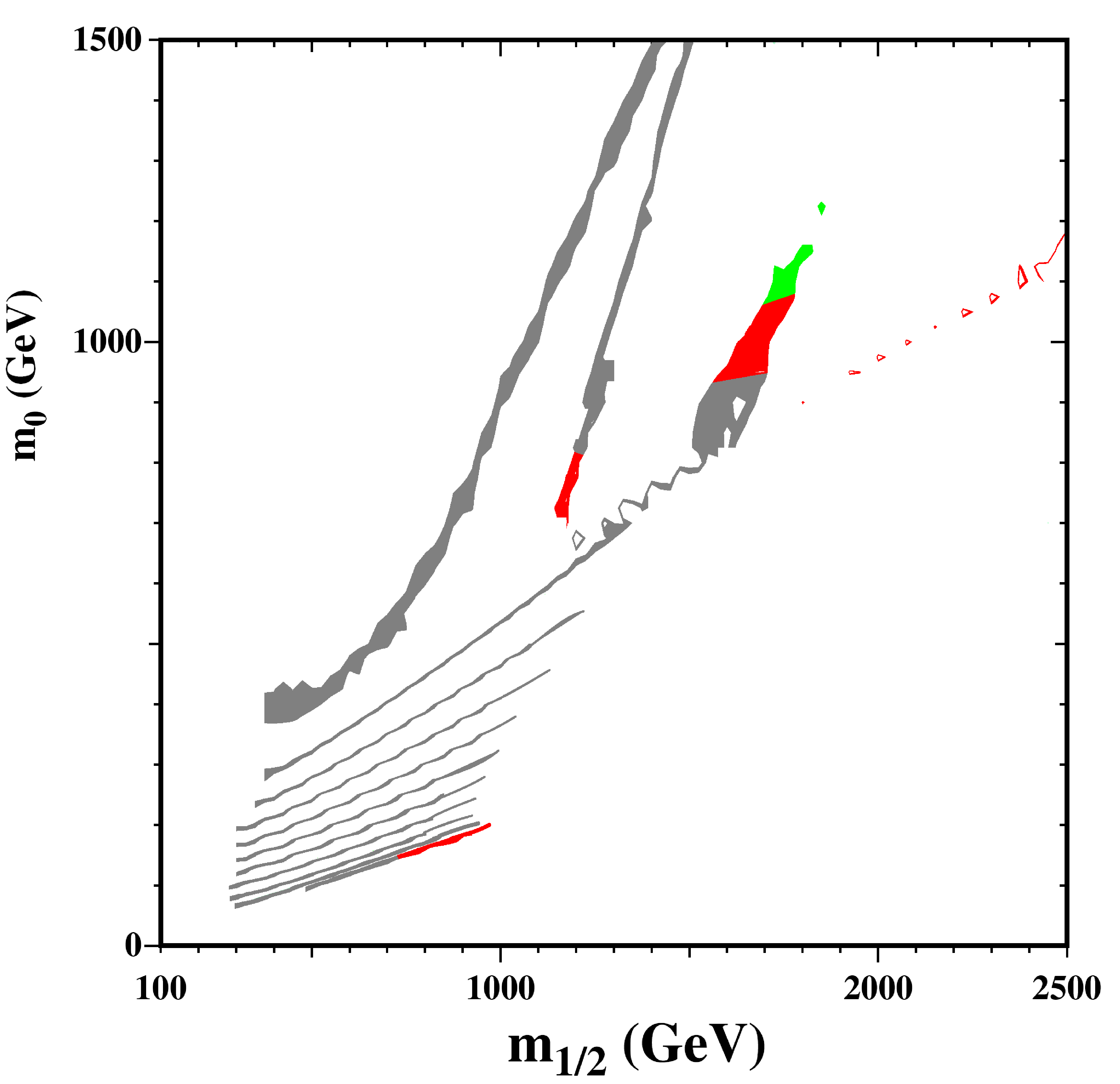}}
   \end{center}
   \caption{{\bf Hierarchical $\nu_R$s}: WMAP and laboratory constraint parameterization of the CMSSM,
   and LFV compliance, based on the expected LFV bound from MEG, BR$(\meg) \lesssim 5\times 10^{-14}$
   for $\tan \beta =5,10,15,20,25,30,35,40,45,50,55$, $\mu>0$ and $a=0$.
   Grey indicates that less than $25$\% of the range of $\theta_1$ is allowed.
   Red indicates that between $25$\% and $50$\% of $\theta_1$ is allowed.
   Green and blue illustrate that $50$\% to $75$\%, and $75$\% to $100$\%, are allowed respectively.
   The constraint regions are reproduced from \cite{Ellis:2003wm}.}
   \label{WMAP2}
\end{figure}

\begin{figure}[ht!]
   \newlength{\picwidthc}
   \setlength{\picwidthc}{7in}
   \begin{center}
       \resizebox{\picwidthc}{!}{\includegraphics{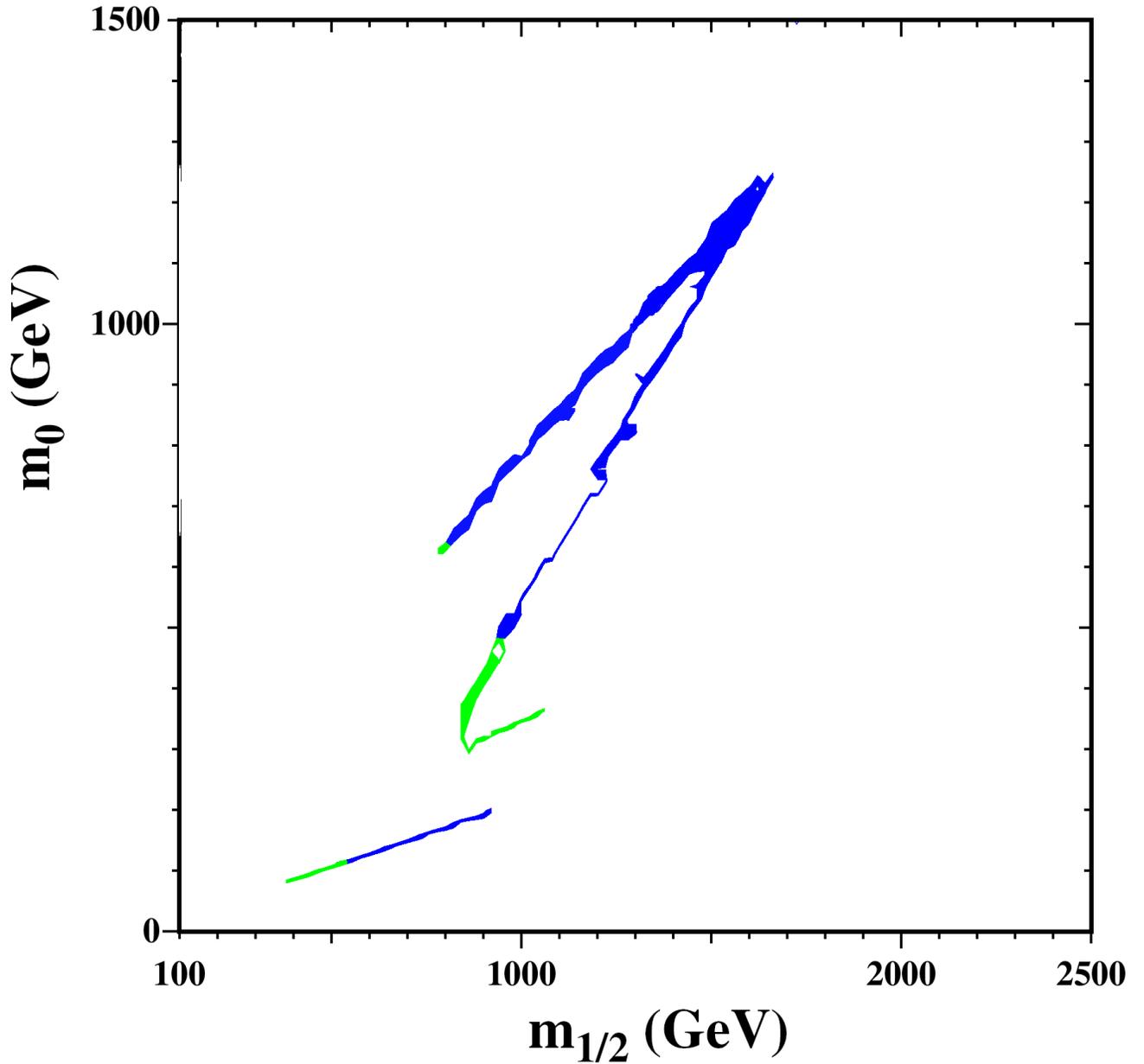}}
   \end{center}
   \caption{{\bf Hierarchical $\nu_R$s}: WMAP and laboratory constraint parameterization of the CMSSM,
   and LFV compliance, based on the current LFV bound, BR$(\meg)<1.2\times 10^{-11}$
   for $\tan \beta =10,35$, $\mu<0$ and $a=0$.
   Grey indicates that less than $25$\% of the range of $\theta_1$ is allowed.
   Red indicates that between $25$\% and $50$\% of $\theta_1$ is allowed.
   Green and blue illustrate that $50$\% to $75$\%, and $75$\% to $100$\%, are allowed respectively.
   The constraint regions are reproduced from \cite{Ellis:2003bm}.
   }
   \label{WMAP3}
\end{figure}

\begin{figure}[ht!]
   \newlength{\picwidthq}
   \setlength{\picwidthq}{7in}
   \begin{center}
       \resizebox{\picwidthq}{!}{\includegraphics{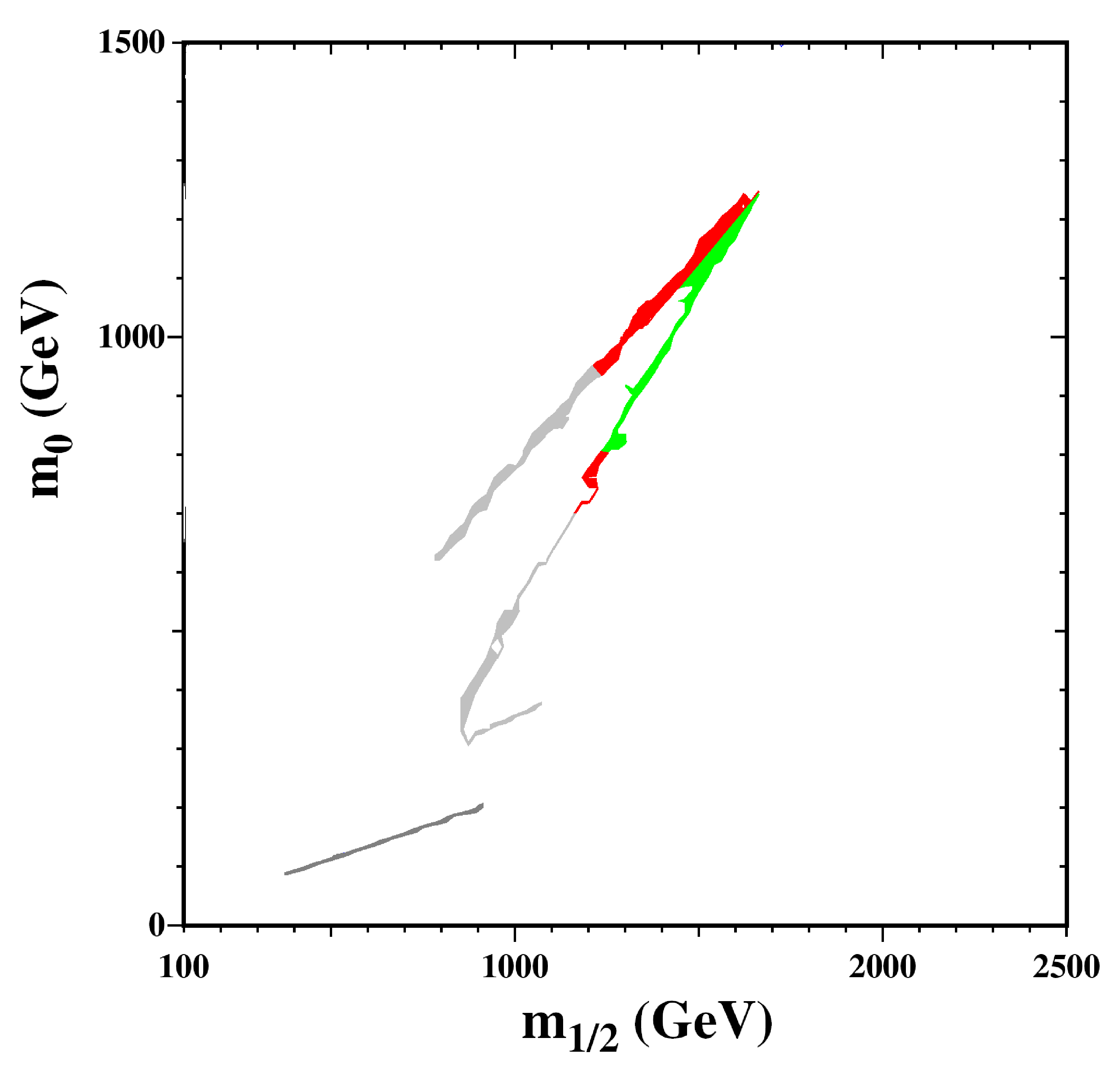}}
   \end{center}
   \caption{{\bf Hierarchical $\nu_R$s}: WMAP and laboratory constraint parameterization of the CMSSM,
   and LFV compliance, based on the expected LFV bound from MEG, BR$(\meg)\lesssim 5\times 10^{-14}$
   for $\tan \beta =10,35$, $\mu<0$ and $a=0$.
   Grey indicates that less than $25$\% of the range of $\theta_1$ is allowed.
   Red indicates that between $25$\% and $50$\% of $\theta_1$ is allowed.
   Green and blue illustrate that $50$\% to $75$\% and $75$\% to $100$\% are allowed respectively.
   The constraint regions are reproduced from \cite{Ellis:2003bm}.
   }
   \label{WMAP4}
\end{figure}

In order to quantify and further illustrate the regions that are both LFV and CMSSM
compliant in this scenario,
consider figures \ref{WMAP1} and \ref{WMAP2}. The regions considered
are a parameterization of the WMAP data from \cite{Ellis:2003wm}.
In figure \ref{WMAP1}, the bands correspond to $\tan \beta =5,10,15,20,25,30,35,40,45,50,55$ for $\mu>0$ and each colour represents
the percentage of the $\theta_1$ range that is allowed by the current bound on $\meg$, $\mathrm{BR}(\meg) < 1.2 \times 10^{-11}$.
Grey indicates that less than $25$\% of $\theta_1$ is allowed, while
red, green and blue illustrate that between $25$\% and $50$\%, $50$\% and $75$\%,
and between $75$\% and $100$\% is allowed respectively. Notice that there are two competing effects controlling the
amount of LFV in these plots. As we move higher in $\tan\beta$, the branching ratio, BR$(\meg)$ increases as eq.(\ref{tanbdepend}).
At the same time, the rate becomes suppressed at larger $m_0$ and $m_{1/2}$. As figure \ref{WMAP1} illustrates,
there are portions of the parameter space at high $\tan\beta$, ($\ie \gtrsim 45$), that are consistent with the current LFV bound
due to the high universal scalar and gaugino mass in those regions. Figure \ref{WMAP2} shows the situation after a possible null
result from MEG, ($\mathrm{BR}(\meg) \lesssim 5\times 10^{-14}$). We see that a large portion of the parameter space would be highly
restricted, with most of the parameter space relegated to less than $25$\%. Therefore, the $\theta_1$ range will be throughly probed by the
up coming experiments, given this see-saw scenario.
In the $\mu<0$ case, the situation is slightly different. While the branching ratio of $\meg$ is largely insensitive to the sign of $\mu$,
the WMAP compliant parameter space is not \cite{Ellis:2003bm}. Figure \ref{WMAP3} shows the constraints from lepton flavour violation
with the current limit on $\mathrm{BR}(\meg) < 1.2 \times 10^{-11}$ over the WMAP range for $\mu<0$ and $a=0$ with $\tan\beta=10,35$.
Grey indicates that less than $25$\% of $\theta_1$ is allowed, while
red, green and blue illustrate that between $25$\% and $50$\%, $50$\% and $75$\%,
and between $75$\% and $100$\% is allowed respectively.
The funnel structure in figure \ref{WMAP3} for $\mu<0$ appears at lower
$\tan \beta$ ($\ie \sim 35$)
compared to figure \ref{WMAP1}.
This pushes the parameter space to larger values of $m_0$ and $m_{1/2}$ at lower $\tan \beta$ and
therefore allows more room where the WMAP region is LFV compliant. Figure \ref{WMAP4} shows how figure \ref{WMAP3} changes after the expected
results from MEG.
If LFV is not observed in the near future, this scenario will only allow a small region of $\theta_1$
corresponding to values near those given in eq.(\ref{sangle}) with $\mu >0$, or a relatively moderate region of $\theta_1$ with $\mu<0$.

\subsection{{Degenerate $\nu_R$s}}

\begin{figure}[ht!]
   \newlength{\picwidthx}
   \setlength{\picwidthx}{7in}
   \begin{center}
       \resizebox{\picwidthx}{!}{\includegraphics{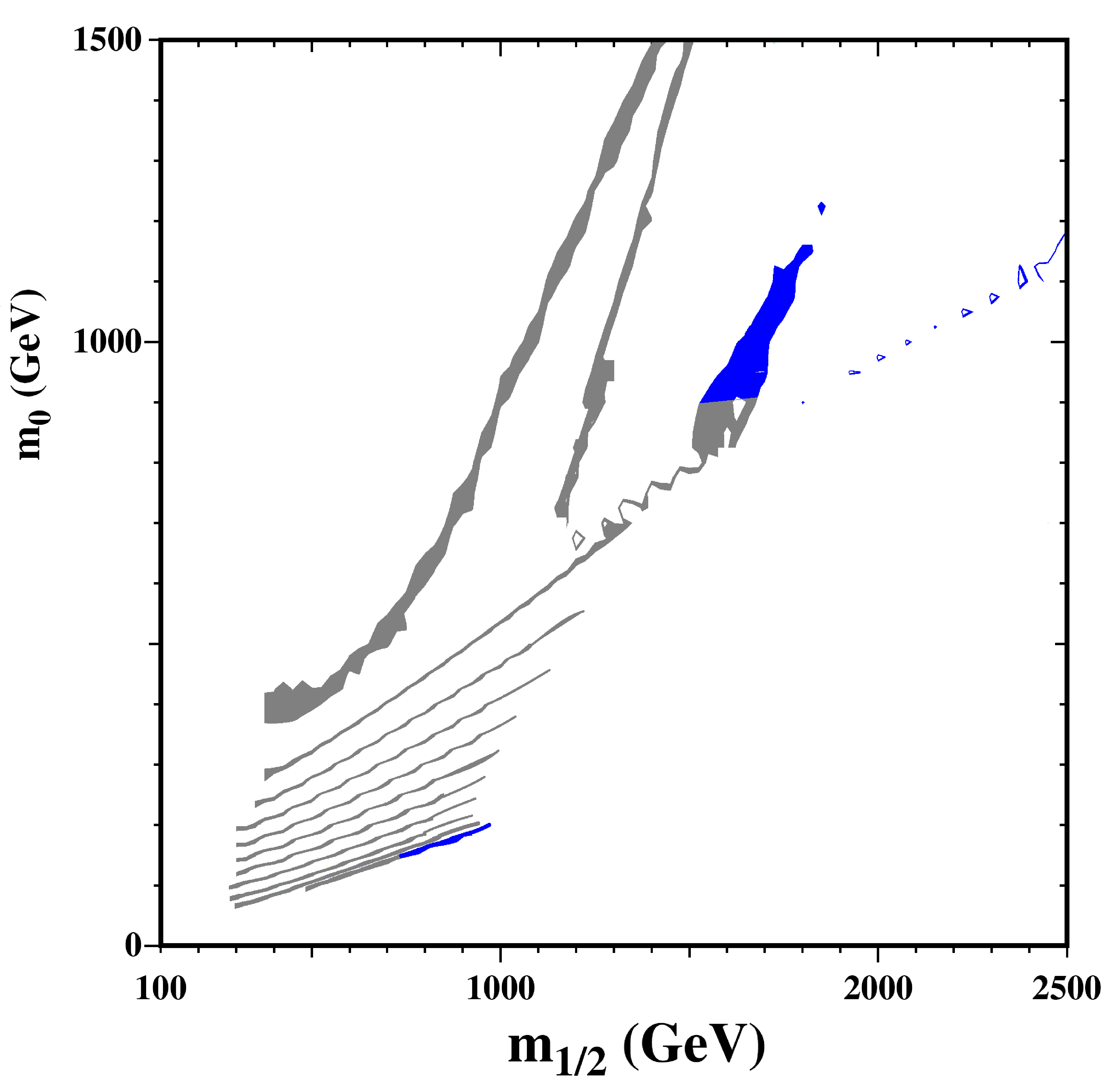}}
   \end{center}
   \caption{{\bf Degenerate $\nu_R$s}: WMAP and laboratory constraint parameterization of the CMSSM,
   and LFV compliance, based on the current LFV bound, BR$(\meg)<1.2\times 10^{-11}$
   for $\tan \beta =5,10,15,20,25,30,35,40,45,50,55$, $\mu>0$ and $a=0$. Blue indicates the allowed region.
   The constraint regions are reproduced from \cite{Ellis:2003wm}.}
   \label{WMAP5a}
\end{figure}

\begin{figure}[ht!]
   \newlength{\picwidthz}
   \setlength{\picwidthz}{7in}
   \begin{center}
       \resizebox{\picwidthz}{!}{\includegraphics{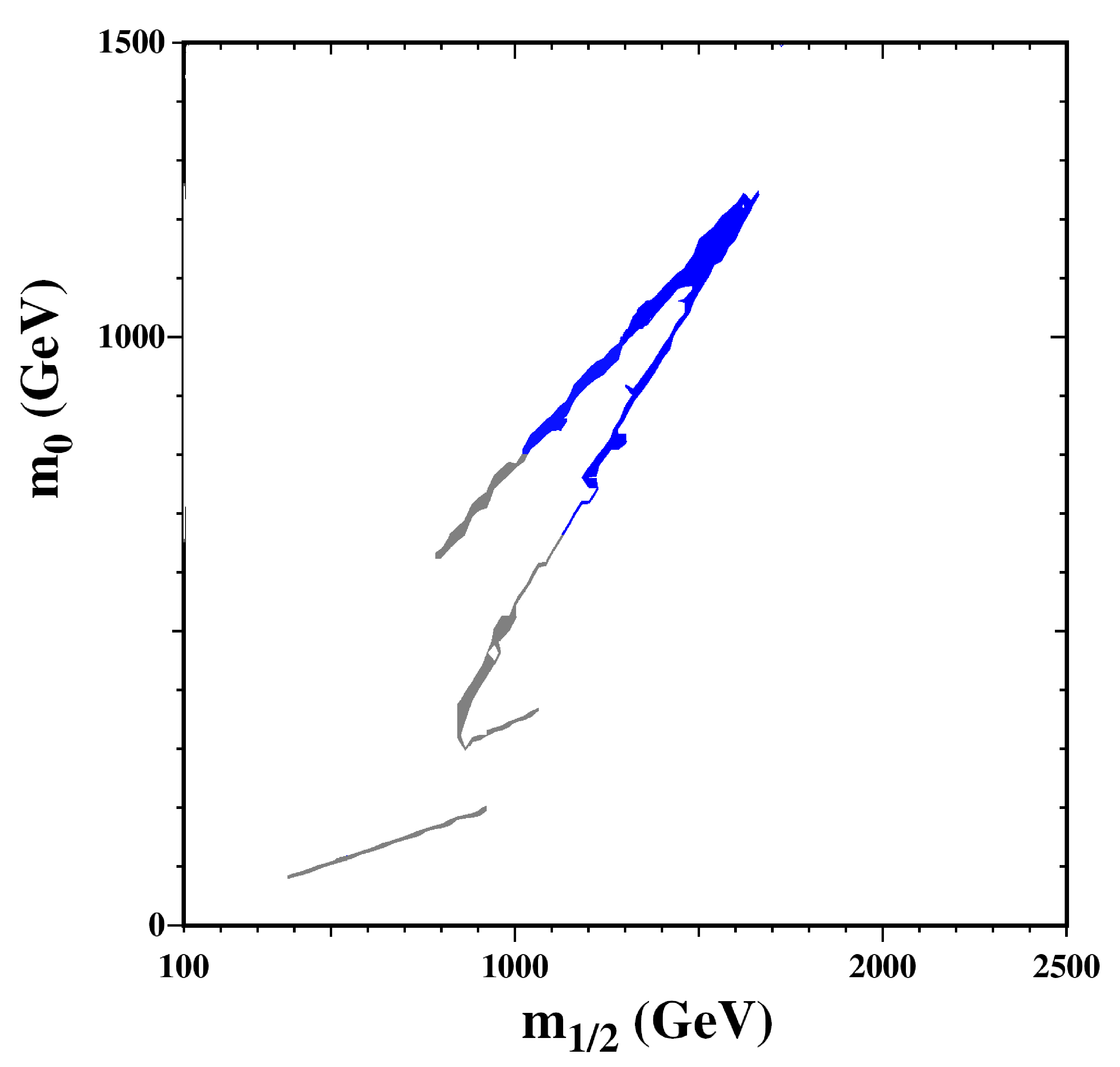}}
   \end{center}
   \caption{{\bf Degenerate $\nu_R$s}: WMAP and laboratory constraint parameterization of the CMSSM,
   and LFV compliance, based on the current LFV bound, BR$(\meg)<1.2\times 10^{-11}$
   for $\tan \beta =10,35$, $\mu<0$ and $a=0$. Blue indicates the allowed region.
   The constraint regions are reproduced from \cite{Ellis:2003bm}.}
   \label{WMAP6a}
\end{figure}

Ignoring possible phases in ${\bf R}$, lepton-flavour violation becomes ${\bf R}$-independent, in the case of degenerate
singlet right-handed neutrino Majorana masses.
We see from eq.(\ref{ynu}) that,
${\bf Y_\nu}{\bf Y_\nu}^\dagger$, which controls the amount of lepton flavour violation
becomes
\be
{\bf Y_\nu}^\dagger {\bf Y_\nu} = {\bf \mathcal{M}} {\bf U} {\bf \kappa}_d {\bf U}^\dagger,
\ee
which is independent of ${\bf R}$. Again, we use the GUT relation $|Y_0|\sim \mathcal{M}
\kappa_3=|Y_t(M_{\mathrm{GUT}})|$ as in \cite{Casas:2001sr}. The situation here is quite different
from the hierarchical case.
Figure \ref{WMAP5a} shows the currently allowed region for $\meg$ consistent
with the CMSSM for $\mu>0$ \cite{Ellis:2003wm}. Notice that most of the parameter space is ruled
out in this scenario; only $\tan \beta \lesssim 5$ and a small region at $\tan\beta \approx 50$ are
consistent with the current LFV bounds.
The upcoming limits will probe all of this currently allowed region.
In the $\mu<0$ \cite{Ellis:2003bm} case more of the parameter space is allowed as the region is pushed to higher soft
mass scales and therefore the LFV rates become suppressed as before. Figure \ref{WMAP6a} illustrates the allowed
region consistent with the current LFV bounds for $\mu<0$.
Clearly the degenerate case, with maximized ``unification" neutrino Yukawa couplings,
is strongly constrained by the present data and will be be severely probed by the forth-coming generation of experiments.


\section{Conclusion}

In this paper, we examined CMSSM lepton-flavour violation in simple general classes of see-saw models \cite{Casas:2001sr}
which had been constructed to fit the data on low energy neutrino oscillations. The models considered have had their neutrino Yukawa
couplings (and Majorana mass scale) chosen as large as reasonable, to maximize the rates for lepton-flavour violating decays. Nevertheless,
when the CMSSM parameters for the models were restricted (following \cite{Ellis:2003wm, Ellis:2003bm}) to have LSP relic
densities in the region determined by WMAP,
and to be consistent with the LEP direct search limits, and the rate for $b \rightarrow s \gamma$, the resulting rate for lepton-flavour
violation was such that over much of the allowed WMAP range, much of the model parameter space was consistent with the present experimental
limit on $\mathrm{BR}(\mu \rightarrow e \gamma)$ (and so, a fortiori, with present limits on the other (charged) lepton-flavour violating processes).
We also noted that the next generation of $\mu \rightarrow e \gamma$ experiments should definitively probe the range of branching ratios suggested
by these models at maximal Yukawa couplings, and also for ranges of smaller Yukawas depending on the CMSSM parameters and the exact see-saw model details.

A future detection of $\mu \rightarrow e \gamma$ would, however, represent not the end of lepton-flavour violation studies of these models, but
rather just the beginning. To disentangle the details of CMSSM see-saw lepton flavour violation will require comparisons of rates for different
LFV muon decays, including $\mu \rightarrow e e e$, and $\mu N \rightarrow e N$. It will also require the observation of (charged) lepton-flavour violation
in different generations, such as $\tau \rightarrow \mu \gamma$,
$\tau \rightarrow e \gamma$, $\tau \rightarrow \mu l l$, and
$\tau \rightarrow e l l$, with $l$ either $e$ or $\mu$. With a combination of observed rates for different LFV $\mu$-decays, and the observation of LFV
in $\tau$ decays, one can hope to begin to uncover both the precise nature of the low-energy soft supersymmetry breaking, as well as the origin of the
lepton-flavour violating interactions responsible for inducing these decays.
Fortunately, we can look forward to a new generation of dedicated
$\mu \rightarrow e \gamma$ \cite{meg}, and $\mu N \rightarrow
e N$ \cite{meco},\cite{prime} experiments, as well as to $\tau$ sources of unprecedented flux, to help us find the experimental signatures of this
new realm of physics.


\section{Acknowledgements}

We are deeply indebted to the authors of \cite{Ellis:2003wm}, and \cite{Ellis:2003bm},
for permission to use their figures for the WMAP and laboratory allowed regions of CMSSM parameter space,
on which we have overlayed our results for the lepton-flavour violation rates. We especially thank Keith Olive for
guidance on the inputs into the determinations of the allowed regions shown in different publications. We would also like to thank
Carlos Wagner and Ernest Jankowski for helpful discussions. BC and DM acknowledge the support of the Natural Sciences and Engineering
Research Council of Canada. BM was supported by the U.S. Department of Energy under contract W-31-109-Eng-38.



\section{Appendix}

\subsection{{$l^-_j\rightarrow l^-_i \gamma$}}

The on-shell amplitude for $l^-_j\rightarrow l^-_i \gamma$ is given
by,
\be
T = e \epsilon^{\alpha} \bar u_i(p-q) \left(m_{l_j}
i \sigma_{\alpha\beta} q^\beta(A^L_2P_L + A^R_2P_R)\right) u_j(p).
\label{Lmeg}
\ee
In the above, $\epsilon$ is the photon polarization vector and
$A^{L,R}_{2}$ are the dipole coefficients.
The decay rate for $\mu \rightarrow e \gamma$ can
be expressed using eq.(\ref{Lmeg}) as,
\be
\Gamma(\mu \rightarrow e \gamma) = \frac{e^2}{16\pi} m^5_{\mu} \left(|A^L_2|^2 + |A^R_2|^2\right).
\ee
Each dipole coefficient can be broken up into the sum of
two terms,
\be
A^{L,R}_{2} = A^{(n)L,R}_{2} + A^{(c)L,R}_{2},
\ee
where $(c)$ and $(n)$ refer to the chargino and neutralino loop
contributions. The expressions \cite{Hisano:1995nq}
for these contributions are,

\bea
A^{(n)L}_2 &=& \frac{1}{32\pi^2} \frac{1}{m^2_{\tilde l_X}}
\left[N^{L(l)}_{iAX}N^{L(l)*}_{jAX} \frac{1}{6(1-r_n)^4} \right. \nonumber \\
&&\times (1-6r_n + 3r^2_n + 2r^3_n -6r_n^2 \ln r_n)\nonumber \\
&&+ \left.
N^{L(l)}_{iAX}N^{R(l)*}_{jAX} \frac{M_{\tilde \chi^0_A}}{m_{l_j}} \frac{1}{(1-r_n)^3}(1-r_n^2 + 2r_n\ln r_n)\right] \\
A^{(n)R}_2 &=& A^{(n)L}_2 |_{L \leftrightarrow R},\\
A^{(c)L}_2 &=& -\frac{1}{32\pi^2} \frac{1}{m^2_{\tilde \nu_X}}\left[C^{L(l)}_{iAX}C^{L(l)*}_{jAX}\frac{1}{6(1-r_c)^4} \right. \nonumber \\
&&\times(2 + 3r_c -6r_c^2 + r_c^3 + 6r_c \ln r_c) \nonumber \\
&& +\left. C^{L(l)}_{iAX}C^{R(l)*}_{jAX} \frac{M_{\tilde\chi^-_A}}{m_{l_j}} \frac{1}{(1-r_c)^3} (-3+4r_c -r_c^2 -2 \ln r_c)\right]\\
A^{(c)R}_2 &=& A^{(c)L}_2 |_{L \leftrightarrow R},
\eea
where $r_n=M_{\tilde \chi^0_A}/m^2_{\tilde l_X}$ and $r_c=M_{\tilde \chi^-_A}/m^2_{\tilde \nu_X}$, and

\bea
C^R_{iAX} &=& -g_2 (O_R)_{A1} U^\nu_{X,i} \\
C^L_{iAX} &=& g_2\frac{m_{l_i}}{\sqrt{2} m_W \cos \beta} (O_L)_{A2} U^\nu_{X,i},
\eea

\bea
N^{R(l)}_{iAX} &=& -\frac{g_2}{\sqrt{2}}\left([-(O_N)_{A2} - (O_N)_{A1} \tan \theta_W]U^l_{X,i} + \frac{m_{l_i}}{m_W \cos\beta} (O_N)_{A3}U^l_{X,i+3}\right)\\
N^{L(l)}_{iAX} &=& -\frac{g_2}{\sqrt{2}}\left(\frac{m_{l_i}}{m_W\cos\beta} (O_N)_{A3}U^l_{X,i} + 2(O_N)_{A1}\tan \theta_W U^l_{X,i+3}\right).
\eea
The matrices in the above expressions are defined in \cite{Hisano:1995nq}.

\subsection{Renormalization Group Equations}

For energy scales above the Majorana scale $\mathcal{M}$, the seesaw sector propagates unsuppressed.
Using the notation, $d\mathbf{X}/d\ln Q = \dot\mathbf{X}/16\pi^2$, the beta functions of the seesaw sectors are:
\begin{equation}
\label{eq-Yn}
\dot\Yn=\Yn\left(
- \gy^2 \unit
- 3 g_2^2 \unit
+ 3\tr\left(\Yu^\dagger \Yu \right) \unit
+ \tr\left(\Yn^\dagger \Yn \right) \unit
+ 3 \Yn^\dagger \Yn
+ \Ye^\dagger \Ye
\right)
\end{equation}
\begin{equation}
\label{eq-mn2}
\dot\mns =
2 \mns \Yn \Yn^\dagger
+ 2 \Yn \Yn^\dagger \mns
+ 4 \Yn \mls \Yn^\dagger
+ 4 \mhus \Yn \Yn^\dagger
+ 4 \An \An^\dagger
\end{equation}
\begin{eqnarray}
\label{eq-An}
\dot\An &=&
-\gy^2 \An
-3 g_2^2 \An
+3 \tr\left( \Yu^\dagger \Yu \right) \An
+ \tr\left( \Yn^\dagger \Yn \right) \An \\
& & - 2 \gy^2 M_1 \Yn
- 6 g_2^2 M_2 \Yn
+ 6 \tr\left( \Yu^\dagger \Au \right) \Yn
+ 2 \tr\left( \Yn^\dagger \An \right) \Yn \nonumber \\
& & + 4 \Yn \Yn^\dagger \An
+ 5 \An \Yn^\dagger \Yn
+ 2 \Yn \Ye^\dagger \Ae
+ \An \Ye^\dagger \Ye \nonumber
\end{eqnarray}
While above the Majorana scale, the beta functions of the CMSSM couplings and masses are augmented by the seesaw sector.
\begin{equation}
\label{eq-Ye}
\Delta\dot\Ye = \Ye\Yn^\dagger \Yn
\end{equation}
\begin{eqnarray}
\label{eq-mhu2}
\Delta \dot\mhus &=&
2 \tr\left(
  \mls \Yn^\dagger \Yn
+ \Yn^\dagger \mns \Yn
+ \mhus \Yn^\dagger \Yn
+ \An^\dagger \An
\right)
\end{eqnarray}
\begin{eqnarray}
\label{eq-ml2}
\Delta\dot\mls &=&
\mls \Yn^\dagger \Yn
+ \Yn^\dagger \Yn \mls \\
& & + 2 \Yn^\dagger \mns \Yn
+ 2 \mhus \Yn^\dagger \Yn
+ 2 \An^\dagger \An
\end{eqnarray}
\begin{eqnarray}
\label{eq-Ae}
\Delta\dot\Ae &=&
\Ye \Yn^\dagger \An
+ \Ae \Yn^\dagger \Yn
\end{eqnarray}


\newpage

\end{document}